\documentclass{article}
\usepackage{arxiv}

\usepackage[utf8]{inputenc} 
\usepackage[T1]{fontenc}    
\usepackage{url}            
\usepackage{booktabs}       
\usepackage{amsfonts}       
\usepackage{nicefrac}       
\usepackage{microtype}      
\usepackage{graphicx}
\usepackage{doi}
\usepackage[all]{xy}
\usepackage{setspace}
\usepackage{xcolor}
\usepackage{tikz}
\usetikzlibrary{cd}
\usepackage{cancel}
\usepackage{caption}
\usepackage{subcaption}

\title{Multiscale heat transport with inertia and thermal vortices}

\author{
Michal Pavelka \\
 Mathematical Institute, Faculty of Mathematics, Charles University,\\ 
 Sokolovsk\'{a} 83, 18675 Prague, Czech Republic
\And 
Liliana Restuccia\\
Department of Mathematical and Computer Sciences, Physical Sciences and Earth Sciences University of Messina,\\
Viale F. Stagno d’Alcontres, Salita Sperone 31, 98166, Messina, Italy
\And David Jou\\
Department of Physics, Universitat Autonoma de Barcelona\\
08193 Bellaterra, Catalonia, Spain
}

\newcommand{\shorttitle}{Heat with vortices}

\usepackage{amsmath,amssymb,amsfonts,latexsym,float,graphics,epsfig}
\usepackage[all]{xy}

\newcommand{\xx}{\mathbf{x}}
\newcommand{\rr}{\mathbf{r}}
\newcommand{\pp}{\mathbf{p}}

\newcommand{\ww}{\mathbf{w}}
\newcommand{\nn}{\mathbf{n}}

\newcommand{\qq}{\mathbf{q}}

\newcommand{\mm}{\mathbf{m}}
\newcommand{\oomega}{\boldsymbol{\omega}}

\newcommand{\LL}{\mathbf{L}}
\newcommand{\KK}{\mathbf{K}}
\newcommand{\MM}{\mathbf{M}}
\newcommand{\rev}{\mathrm{rev}}
\newcommand{\irr}{\mathrm{irr}}
\newcommand{\Par}{\mathcal{P}}
\newcommand{\OBig}{\mathcal{O}}
\newcommand{\vv}{\mathbf{v}}
\newcommand{\VV}{\mathbf{V}}

\begin{document}
\maketitle

\begin{abstract}
In this paper, we present a Hamiltonian and thermodynamic theory of heat transport on various levels of description. Transport of heat is formulated within kinetic theory of polarized phonons, kinetic theory of unpolarized phonons, hydrodynamics of polarized phonons, and hydrodynamics of unpolarized phonons. These various levels of description are linked by Poisson reductions, where no linearizations are made. Consequently, we obtain a new phonon hydrodynamics that contains convective terms dependent on vorticity of the heat flux, which are missing in the standard theories of phonon hydrodynamics. Moreover, the equations are hyperbolic and Galilean invariant, unlike current theories for beyond-Fourier heat transport. The vorticity-dependent terms violate the alignment of the heat flux with the temperature gradient even in the stationary state, which is expressed by a Fourier-Crocco equation. The new terms also cause that temperature plays in heat transport a similar role as pressure in aerodynamics. 
\end{abstract}

\tableofcontents

\doublespace
\section{Introduction}
This manuscript contains a new theory of phonon hydrodynamics with corrected behavior of thermal vortices. Phonon hydrodynamics is of particular importance for instance in nanoscale heat transport in semiconductors, thermal diodes, or two-dimensional materials  \cite{guo2015,torres2018,sendra2021}. 

\begin{figure}[ht!]
\begin{equation*}
\xymatrix{f_{T+},f_{T-},f_L \ar[rr]\ar[dd]&& \mm_{T+},\mm_{T-},\mm_{T_L},s_{T+},s_{T-},s_{T_L}    \ar[dd]\\ \\  f  \ar[rr] && s,\mm}
\end{equation*}
\caption{
Distribution functions of positively and negatively transversely and longitudinally polarized phonons, $f_{T+},f_{T-},f_L$, can be summed to the overall distribution function $f$. They can be also projected to the momentum and entropy densities of the particular polarizations, $\mm_{T+}$, $\mm_{T-}$, $\mm_{T_L}$, $s_{T+}$, $s_{T-}$, and $s_{T_L}$. These densities can be then summed to the overall momentum density and entropy density of phonons, $\mm$ and $s$. Finally, the overall distribution function $f$ can be projected to $\mm$ and $s$. The projection of the Poisson bracket of kinetic theory to the hydrodynamic fields then leads to a new theory of phonon hydrodynamics, similarly as in \cite{gheat}.}
\label{fig.diagram}
\end{figure}
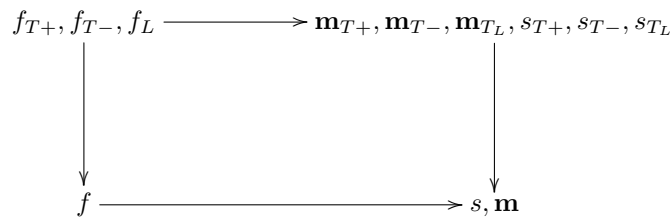

Heat transport in solids is represented by transport of energy by lattice vibrations \cite{kittel,peiersl}. Since the lattice vibrations can be seen as quasiparticles obeying Hamilton canonical equations (phonons), evolution equation of their distribution function obeys kinetic theory \cite{peierls-kinetic,mr}, which is also a Hamiltonian system \cite{mar82,grcontmath}. Here, we start with kinetic theory of phonons with three polarizations (two transverse and one longitudinal). This description is then reduced to hydrodynamics with momenta and entropies of the particular polarizations, and to the overall kinetic theory of phonons (disregarding polarizations). Both these levels of description are then further reduced to a new theory of phonon hydrodynamics, where state variables are the overall momentum density and entropy density. Diagram \ref{fig.diagram} summarizes the various levels of description and connections between them.

A particular feature of our phonon hydrodynamics is that the evolution equation for the conjugate entropy flux $\ww=\mm/s$ (the ratio of phonon momentum and entropy, see Equation \eqref{eq.w}) contains terms depending on the thermal vorticity, $\oomega = \nabla \times \ww$. In contrast with other approaches, our theory consistently describes the motion of thermal vorticity because we neither linearise the reversible part of the equations nor are we restricted to one-dimensional systems \cite{guo2015,guyer-krumhansl,zhang-vortices,Kovacs-Van,struch-dreyer}. The terms that depend on $\oomega$ make the equations Galilean invariant, unlike the usual Maxwell-Cattaneo-Vernotte or Guyer-Krumhansl equations \cite{catt,guyer-krumhansl}. Although evidence for the presence of thermal vorticity in heat transport is provided by kinetic theory \cite{shang2020, Zhang2021,raya-moreno}, standard phonon hydrodynamics can not correctly propagate such vortices because it lack convective terms. Moreover, the missing convective terms are the reason why the standard theories of phonon hydrodynamica violate Galilean invariance even when all collision terms are neglected. The convective terms are lost in the standard linearized Chapman-Enskog reduction, while our method (based on reduction the underlying Poisson brackets) does not linearize the reversible evolution and recovers the convective terms. In other words, phonon hydrodynamics presented in this manuscript corrects the standard phonon hydrodynamics so that heat vortices are transported in agreement with the kinetic theory.

Moreover, the new terms cause that the heat flux no longer needs to be aligned with the gradient of temperature even in the stationary state. Instead, the stationary state is described by a Fourier-Crocco equation \eqref{eq.crocco}, where the heat flux, thermal vorticity, and temperature gradient can form a triplet of mutually perpendicular vectors in the limiting case of negligible dissipation. Our equations also represent a system of quasilinear first-order hyperbolic equations, so it describes hyperbolic heat conduction \cite{shtc-generic,God-Siberian,Romensky-hyperbolic,ader-vis} (if we neglect the viscous-like dissipative terms). In the one-dimensional approximation and after linearization, our theory simplifies to the standard Maxwell-Cattaneo-Vernotte equation \cite{catt}. Finally, when we go beyond the zero-th Chapman-Enskog approximation in the irreversible evolution of phonon momentum density, we obtain higher-order (Laplacian) terms that correspond to the dissipation in the Guyer-Krumhansl equation. Therefore, our phonon hydrodynamics can be seen as a non-linear generalization of the (linear) Guyer-Krumhansl equation \cite{guyer-krumhansl}. Dynamics of thermal vorticity based on solutions of the phonon kinetic theory was already observed in \cite{Zhang2021,raya-moreno}, but here we also formulate the corresponding phonon hydrodynamics.

A theory of hyperbolic heat transport was developed in the context of Extended Rational Thermodynamics (RET) \cite{ruggeri-strumia,struch-dreyer,mr}, including comparison with the heat-pulse experiment \cite{naf}. In the present manuscript, we show the Hamiltonian structure of the theory without linearization, which keeps also the non-linear vorticity-dependent terms, that are important for inertial effects, for effects of thermal vorticity, and for Galilean invariance. 

Another description of hyperbolic heat transport was developed within the Internal Variables Theory (IVT) \cite{Kovacs-Van,van-berezovski}. This theory is not constructed by reduction from a more microscopic description (in contrast with kinetic theory), but by extending the Fourier heat equation by including additional internal variables and parameters, which remain unspecified. IVT shows a better agreement with the experimental data than the kinetic theory for the same number of state variables, but it also includes more fitting parameters. Comparison of IVT with the GENERIC framework (used in this paper) can be found in \cite{Matyas2021}.

Apart from RET as IVT, there are other theories of hyperbolic heat transport coming from non-equilibrium thermodynamics (NET). NET was actually founded by Boltzmann \cite{boltzmann}, Onsager \cite{onsager1930,onsager1931}, and others \cite{meixner,dgm,eckart}. Subsequent theories of NET, as Rational (Extended) Thermodynamics \cite{trus,mr}, or Extended Irreversible Thermodynamics \cite{david-eit,david-liliana}, bring the possibility to derive evolution equations for additional state variables. There is plenty of successful applications of these theories, ranging from non-Newtonian fluids and solids to theories of charged mixtures \cite{lebon-understanding}. However, it is often difficult to fully determine the reversible terms that are not visible in the formula for entropy production, which is why we use a different framework, that builds upon Hamiltonian mechanics and has already been applied to radiation hydrodynamics \cite{miroslav-radiation}. Here, we rely on the General Equation for Non-Equilibrium Reversible-Irreversible Coupling (GENERIC) \cite{go,og,hco,pkg}, which combines Hamiltonian continuum mechanics and with gradient dynamics.

The plan of this paper follows. Section \ref{sec.kinetic.pol} recalls kinetic theory of polarized phonons, in particular the evolution of the distribution function of the phonon quasiparticles. Section \ref{sec.kin} then simplifies the description to the overall distribution of phonons, disregarding the polarizations, and it adds irreversible evolution in the form of the Callaway model (collisions of phonons between themselves and with impurities of the lattice). Section \ref{sec.hydro} represents the main part of this paper as it contains the phonon hydrodynamics that emerges by reduction of the kinetic theory. The reversible part of the hydrodynamic theory is expressed by Equations \eqref{eq.m.rev} or \eqref{eq.Cat.s} while their full form (in the zero-th Chapman-Enskog approximation) is in Equations \eqref{eq.Cat.final}. Equations \eqref{eq.me.full} then show phonon hydrodynamics within the first Chapman-Enskog approximation, where higher-order dissipative terms (Laplacian-like) appear. The Fourier-Crocco equation \eqref{eq.crocco}relates thermal vorticity, temperature gradient, and heat flux, which leads to a similarity between temperature in heat transfer and pressure in aerodynamics (Figure \ref{fig.aero}). In order to obtain the enhanced precision, we rely on geometric mechanics and Poisson reductions, which represent a cornerstone of the GENERIC framework, summarized in the following Section.

\subsection{Introduction to GENERIC}
The General Equation for Non-Equilibrium Reversible-Irreversible Coupling (GENERIC) framework combines (reversible) Hamiltonian evolution with (irreversible) gradient dynamics. The framework was established in \cite{go,og} and further developments can be found in monographs \cite{hco,pkg}.\footnote{It is equivalent with the metriplectic systems if the dissipation potential is considered quadratic \cite{mor}. However, in cases with non-quadratic dissipation potentials, both frameworks are different \cite{miroslav-guide}. The GENERIC framework has also a precursor in the so called one-generator formalism \cite{be} and in the nonlinear Onsager-Casimir (NOC) equations \cite{driven}.} An advantage of GENERIC is that it determines the reversible part of the evolution of given state variables because it better exploits the results of geometrical mechanics. In the rest of this Section, we shall recall the GENERIC framework.

Let $\xx$ be a set of state variables and let $A(\xx)$ be a functional on the space (or manifold) where $\xx$ plays the role of coordinates. Evolution of that functional is then given within GENERIC in two equivalent forms,
\begin{equation}\label{eq.generic}
\dot{x}^i = L^{ij} \frac{\delta E}{\delta x^j} + \frac{\delta \Xi}{\delta x^*_i}\Big|_{\xx^* = \frac{\delta S}{\delta \xx}}
    \quad\text{or}\quad
\dot{A} = \{A,E\} + \left\langle \frac{\delta A}{\delta x^i}, \frac{\delta \Xi}{\delta x^*_i}\right\rangle\big|_{\xx^* = \frac{\delta S}{\delta \xx}},
\end{equation}
where the latter is a sort of weak formulation. The evolution consists of two parts, a reversible one and an irreversible one. Note that $\delta \bullet/\delta \bullet$ is the functional derivative, see \cite{hco,pkg,be}, $\langle\bullet,\bullet\rangle$ is a duality (also represented by contraction over repeated indexes), $E$ is the total energy of the isolated system under consideration, $\Xi(\xx,\xx^*)$ is a dissipation potential, and $S$ is the total entropy. We denote by $\xx^*$ the dual variables to the state variables $\xx$, which are also called conjugate entropic variables, eventually identified with derivatives of the total entropy $S$.

If the dissipation potential is quadratic, 
\begin{equation}\label{eq.Xi.quad}
\Xi = \frac{1}{2}\left\langle x^*_i, M^{ij}(\xx), x^*_j\right\rangle,
\end{equation}
the GENERIC evolution \eqref{eq.generic} can be rewritten as
\begin{equation}
    \partial_t x^i = L^{ij}E_{x^j} + M^{ij} S_{x^j},
\end{equation}
where the symmetric operator $M^{ij}$ (called a dissipative matrix) works as in the Ginzburg-Landau theory \cite{landau-ginzburg,hutter2013}.

The reversible part is generated by a Poisson bivector $\LL$ or a Poisson bracket $\{\bullet,\bullet\}$, which are related through
\begin{equation}\label{eq.PB.L}
\{A,B\} = \left \langle d A, L^{ij}, dB\right\rangle
= \frac{\delta A}{\delta x^i} L^{ij} \frac{\delta B}{\delta x^j}
    \quad\text{and}\quad
L^{ij} = \{x^i, x^j\}.
\end{equation}
The bivector is a twice contravariant antisymmetric tensor field on the space of the state variables, which is Lie-dragged by the Hamiltonian evolution \cite{fecko}. The latter property then ensures validity of the Jacobi identity,
\begin{equation}
\{A,\{B,C\}\}+
\{B,\{C,A\}\}+
\{C,\{A,B\}\} = 0 \quad \forall A,B,C
\end{equation}
while the antisymmetry of $\LL$ gives antisymmetry of the bracket,
\begin{equation}
\{A,B\} = -\{B,A\} \quad\forall A,B.
\end{equation}
A particular implication of the antisymmetry is conservation of energy, $\dot{E} = \{E,E\} = 0$, by the Hamiltonian mechanics. Finally, an equivalent way to express the reversible part of the evolution is 
\begin{equation}\label{eq.xrev}
(\partial_t \xx)_\rev = \{\xx,E\},
\end{equation}
as can be verified by the first relation of \eqref{eq.PB.L}. Once we have recalled the GENERIC framework, we are now in position to formulate kinetic theory of phonons within that framework, which is the subject of the following Section.

\section{Kinetic theory of phonons}\label{sec.kinetic.pol}
Phonons are quasiparticles obtained by diagonalization of the Hamiltonian of a crystal \cite{landau3}, and their motion is described by Hamilton canonical equations \cite{kittel}, constructed from the canonical Poisson bracket and from the energy of phonons. The purpose of this Section is to formulate Hamiltonian kinetic theory that governs the distribution functions of the phonons. Actually, phonons have three possible polarizations, two transverse ($T+$ and $T-$) and one longitudinal. Therefore, there are three distribution functions $f_{T+}(t,\rr,\pp)$, $f_{T-}(t,\rr,\pp)$, and $f_{L}(t,\rr,\pp)$. 

\subsection{Kinetic theory of phonons with polarization}
Similarly as the evolution of the distribution function of classical particles is Hamiltonian, and thus generated by a Poisson bracket and energy,  evolution of the phononic distribution functions is also Hamiltonian, and generated by the sum of the Boltzmann Poisson brackets
\begin{equation}\label{eq.PB.fa}
\{A,B\}^{(f_{T+,T-,L})} = \sum_{\alpha\in \{T+,T-,L\}}\int d\rr \int d\pp f_\alpha \cdot \left(\frac{\partial A_{f_\alpha}}{\partial \rr} \cdot \frac{\partial B_{f_\alpha}}{\partial \pp}-\frac{\partial B_{f_\alpha}}{\partial \rr} \cdot \frac{\partial A_{f_\alpha}}{\partial \pp}\right),
\end{equation}
where $A$ and $B$ are two arbitrary functionals of the distribution functions \cite{grpd,pkg}. Symbols $A_{f_\alpha}$ stands for the functional derivative of $A$ with respect to $f_\alpha$.
The reversible evolution equations implied by this Poisson bracket are (using Equation \eqref{eq.xrev})
\begin{equation}
(\partial_t f_\alpha)_{\rev} = -\frac{\partial f_\alpha}{\partial \rr} \cdot \frac{\partial E_{f_\alpha}}{\partial \pp} +\frac{\partial f_\alpha}{\partial \pp} \cdot \frac{\partial E_{f_\alpha}}{\partial \rr} \quad\forall \alpha\in\{T+,T-,L\}.
\end{equation}

What is the energy $E(f_{T+},f_{T-},f_L)$? In general, we need the whole band structure of the crystal to find the precise formula for energy \cite{kittel}. However, if we only restrict to low temperatures, to the first Brillouin zone, and acoustic phonons, we can approximate the energy as
\begin{equation}
E = \int d\rr\int d\pp c|\pp| (f_{T+}+f_{T-}+f_L),
\end{equation}
where $c$ is the Debye velocity and $|\pp|$ is the Eulerian norm of the momentum \cite{debye}. 

The evolution equations obtained from a Poisson bracket are reversible both in the sense of the time-reversal transformation \cite{pre15} and in the sense that they preserve the bosonic entropy \cite{landau5}, 
\begin{multline}\label{eq.Sfa}
S^{(f_{T+,T-,L})} = \sum_\alpha S^{\text{kinetic},\alpha}(f_\alpha)\\
\quad\text{with}\quad
S^{\text{kinetic},\alpha}(f_\alpha) = -\frac{k_B}{h^3}\int d\rr \int d\pp \left(h^3 f_\alpha \ln(h^3 f_\alpha)-(1+h^3f_\alpha)\ln(1+h^3f_\alpha)\right).
\end{multline}
Indeed, any integral of a function of the distribution functions is a Casimir of the Poisson bracket, that is $\{S,B\}^{(f_{T+,T-,L})}=0$ $\forall B$, which can be used also to stability analysis of the Hamiltonian equations \cite{holm-stability}.

Instead of the distribution functions, which contain a large amount of information, we can reduce the description to the hydrodynamic fields of the distribution functions, namely the momentum densities and entropy densities
\begin{subequations}
\begin{align}
\mm_\alpha &= \int d\pp f_\alpha \pp\\
s_\alpha &= \int d\pp \eta_\alpha(f_{\alpha}),
\end{align}
where $\eta_\alpha$ stands for a real-valued function, for instance the integrands in entropy \eqref{eq.Sfa}.
\end{subequations}

The Poisson bracket for the hydrodynamic fields is obtained by plugging functionals that depend only on the hydrodynamic fields, $A(\mm_\alpha(f_\alpha),s(f_\alpha))$, to the Poisson bracket \eqref{eq.PB.fa}. The resulting Poisson bracket reads
\begin{equation}
\{A,B\}^{(\mm_\alpha,s_\alpha)} = \sum_\alpha \int d\rr s_\alpha\left(\partial_i A_{s_\alpha} B_{m_{\alpha i}} - \partial_i B_{s_\alpha} A_{m_{\alpha i}}\right)
+ \sum_\alpha \int d\rr m_{\alpha i}\left(\partial_j A_{m_{\alpha i}} B_{m_{\alpha j}} - \partial_j B_{m_{\alpha i}} A_{m_{\alpha j}}\right),
\end{equation}
and the implied reversible evolution equations are
\begin{subequations}
\begin{align}
\partial_t s_\alpha &= -\partial_j (s_\alpha E_{m_{\alpha j}})\\
\partial_t m_{\alpha i} &= -\partial_j (m_{\alpha i} E_{m_{\alpha j}}) - s_\alpha \partial_i E_{s_{\alpha}} - m_{\alpha j}\partial_i E_{m_{\alpha j}}.
\end{align}
\end{subequations}

The dependence of energy on the hydrodynamic fields can be determined by inverting the formula for entropy of the fields \eqref{eq.SHP}, which was obtained by the principle of maximization of entropy (MaxEnt) in Appendix \ref{sec.maxent}.

Although now we could add a dissipation potential in order to add also an irreversible evolution of the phonons, we shall first reduce the description to a less detailed one, parametrized only by the overall distribution function $f=\sum_\alpha f_\alpha$, which is the subject of the following Section.

\subsection{Overall kinetic theory of phonons}\label{sec.kin}
When the various polarizations of phonons are not important, we may reduce the kinetic theory of polarized phonons to the overall kinetic theory that describes evolution of the overall distribution function 
\begin{equation}\label{eq.f.fa}
f = f_{T+}+f_{T-}+f_L.
\end{equation}

The Poisson bracket for the overall distribution function $f$ follows from the Poisson bracket for state variables $f_{T+}$, $f_{T-}$, and $f_L$ (Equation \eqref{eq.PB.fa}). When we plug functionals that depend only on $f(f_{T+},f_{T-},f_L)$ to the Poisson bracket, we obtain bracket
\begin{equation}\label{eq.PB.B}
\{A,B\}^{(f)} = \int d\rr\int d\pp f \cdot \left(\frac{\partial A_f}{\partial \rr}\frac{\partial B_f}{\partial \pp}-\frac{\partial B_f}{\partial \rr}\frac{\partial A_f}{\partial \pp}\right),
\end{equation}
which is the standard Poisson bracket for the one-particle distribution functions (also called Boltzmann bracket), see for instance \cite{adv,pkg}.
The reversible evolution equation implied by this Poisson bracket is
\begin{equation}\label{eq.f.rev}
(\partial_t f)_{\rev} = -\frac{\partial f}{\partial \rr} \cdot \frac{\partial E_f}{\partial \pp} +\frac{\partial f}{\partial \pp} \cdot \frac{\partial E_f}{\partial \rr}.
\end{equation}

In order to write the evolution equation in a closed form, we need the energy $E(f)$. In our simplified situations (restriction to the first Brillouin zone, acoustic phonons with small $\mathbf{k}$ vectors), the energy is 
\begin{equation}
E = \int d\rr \int d\pp c |\pp| f,
\end{equation}
where $c$ is the Debye speed \cite{kittel}. Once the Debye velocity $c$ is determined, the reversible part of the evolution equation for $f$ is fully specified. 
In particular, the constant $c$ can be obtained from measurement of the second sound speed. 
For instance, for a NaF crystal,
Equation \eqref{eq.secondsound} relates the Debye velocity with the speed of second sound, theoretical limit of which at $T=0$ is $c=2020.9\,m/s$ \cite{hardy}.
 The heat capacity of low-temperature NaF is $c_V = 234 N_A k_B (T/\theta_D)^3$ for temperature between $0.6 K$ and $15 K$; the Debye temperature is found to be $\theta_D = (466\pm 5)K$ \cite{harrison1968}. The density is approximated by $\rho\approx 2.851 g/cm^3$ \cite{birch1979}, and the molar mass is $M\approx 42 g/mol$. 

However, the evolution of $f$ actually contains two parts, a reversible (Hamiltonian) part, which is given by the Poisson bracket and energy, and an irreversible part, which is given by a dissipation potential and entropy. The entropy can be derived from the entropy of phonons with polarization, Equation \eqref{eq.Sfa}. By maximization of the entropy subject to the constraint \eqref{eq.f.fa}, we obtain the entropy
\begin{equation}\label{eq.Sf}
S^{(f)} = -\frac{3k_B}{h^3} \int d\rr \int d\pp \left(h^3 \frac{f}{3}\ln \left(h^3 \frac{f}{3}\right) - \left(1+h^3\frac{f}{3}\right)\ln\left(1+h^3\frac{f}{3}\right)\right).
\end{equation}

The irreversible evolution of the distribution function consists of collisions between phonons (normal process, conserving both energy and momentum) and collisions with impurities of the crystal (resistive process, conserves energy, but not momentum) \cite{mr}. These two processes are described by the so called Callaway model \cite{calaway}
\begin{equation}\label{eq.Callaway}
(\partial_t f)_{\irr} = - \frac{1}{\tau_R}(f - f_{R}) - \frac{1}{\tau_N}(f - f_{N}),
\end{equation}
where $f_R$ and $f_N$ are the MaxEnt estimates of the distribution function obtained by maximization of entropy \eqref{eq.Sf} subject to the constraints given by the overall hydrodynamic momentum and energy \eqref{eq.f.me}, or by the constraint that only the hydrodynamic energy is known, respectively,
\begin{subequations}
\begin{align}
f_N &= \frac{3}{h^3}\frac{1}{e^{\mm^*\cdot\pp/k_B}e^{e^*c|\pp|/k_B}-1}\\
f_R &= \frac{3}{h^3}\frac{1}{e^{e^*c|\pp|/k_B}-1}.
\end{align}
Here, $\mm^*$ and $e^*$ are the derivatives of the hydrodynamic entropy \eqref{eq.sH} with respect to the momentum density and energy density, respectively, 
 $\mm^* = \frac{\partial s^H}{\partial \mm}$ and $e^* = \frac{\partial s^H}{\partial e}$.
\end{subequations}
A feature of distribution functions $f_N$ and $f_R$ is that the Callaway irreversible evolution conserves momentum and energy (in the normal processes) or only the energy (in the resistive processes),
\begin{align}
\int d\pp p_i (f-f_N) = 0, 
\int d\pp |\pp| (f-f_N) = 0, 
\int d\pp |\pp| (f-f_R) = 0.
\end{align}
The normal and resistive relaxation times, $\tau_N$ and $\tau_R$, have typically values around $\tau_R\approx 3 \cdot 10^{-6}s$ and $\tau_N \approx 2 \cdot 10^{-7}s$ \cite{struch-dreyer}.


The final evolution equation is the sum of the reversible part \eqref{eq.f.rev} and irreversible part \eqref{eq.Callaway},
\begin{equation}\label{eq.f}
\partial_t f = -\frac{\partial f}{\partial \rr} \cdot \frac{\partial E_f}{\partial \pp} +\frac{\partial f}{\partial \pp} \cdot \frac{\partial E_f}{\partial \rr}
 - \frac{1}{\tau_R}(f - f_{R}) - \frac{1}{\tau_N}(f - f_{N}).
\end{equation}
The next Section contains a further reduction of this kinetic theory to phonon hydrodynamics.

\section{Overall hydrodynamics of phonons}\label{sec.hydro}
Let us now further reduce the Boltzmann Poisson bracket, that expresses kinematics of kinetic theory, to the Poisson bracket expressing kinematics of the hydrodynamic fields of momentum density and entropy density,
\begin{subequations}\label{eq.us}
\begin{align}
m_i(t,\rr) &= \int d\pp p_i f(t,\rr,\pp)\\
s(t,\rr) &= \int d\pp \eta(f(t,\rr,\pp)).
\end{align}
Here, $\eta(f)$ is the entropy density of phonons $s^H$ (Equation \eqref{eq.sH}), but in general it can be an arbitrary function of $f$. 
\end{subequations}

\subsection{Reversible evolution}
Plugging arbitrary functionals $A$ and $B$  dependent only on momentum and entropy densities, $A(\mm,s)$ and $B(\mm,s)$, into bracket \eqref{eq.PB.B}, we obtain a part of the Poisson brackets of fluid mechanics, 
\begin{equation}
\{A,B\}^{(\mm,s)} = \int d\rr s\left(\partial_i A_s B_{m_i} - \partial_i B_s A_{m_i}\right)
+ \int d\rr m_i\left(\partial_j A_{m_i} B_{m_j} - \partial_j B_{m_i} A_{m_j}\right),
\end{equation}
which is actually the same as the part of the Poisson bracket for fluid mechanics of normal particles (without mass density), see \cite{gheat,pkg} for details of the calculation. The reversible evolution implied by this bracket reads
\begin{subequations}
\begin{align}
\partial_t s &= -\partial_j (s E_{m_{j}})\\
\label{eq.m.rev}
\partial_t m_{i} &= -\partial_j (m_{i} E_{m_{j}}) - s \partial_i E_{s} - m_{j}\partial_i E_{m_{i}}.
\end{align}
\end{subequations}
Similarly as in fluid mechanics, the momentum equation implied by this Poisson bracket contains a non-linear convective term, which expresses inertial effects of the motion of phonons and which is missing in the usual models of hyperbolic heat conduction as the linear Guyer-Krumhansl model \cite{guyer-krumhansl,guo2015}. Although the heat flux is proportional to the momentum density (Equation \eqref{eq.qm}), the convective terms can not be written as $\qq\cdot\nabla\qq$, since the derivative $E_\mm$ also depends on the field of energy density (see Equation \eqref{eq.m.ch1.approx}).

It is advantageous to transform the momentum density to the field 
\begin{equation}\label{eq.w}
\ww = \mm/s
\end{equation}
because then the subsequent reduction to the Fourier heat conduction becomes simpler, see Section \ref{sec.F}. Derivatives of functionals of $\mm$ and $s$ transform as
\begin{equation}
    F_{\mm} \rightarrow \frac{1}{s}F_{\ww}
    \quad\mbox{and}\quad
    F_{s} \rightarrow F_{s} - \frac{1}{s}F_{\ww}\cdot\ww,
\end{equation}
and when we plug functionals dependent only on $(s,\ww)$ into the Poisson bracket for $(\mm,s)$, we obtain a Poisson bracket for $s$ and $\ww$,
\begin{equation}\label{eq.PB.Cat}
\{A,B\}^{(Cattaneo)} = \int d\rr \left(\partial_i A_s B_{w_i}-\partial_i B_s A_{w_i}\right) 
+ \int d\rr \frac{1}{s}(\partial_i w_j - \partial_j w_i)A_{w_i} B_{w_j}.
\end{equation}
We refer to this Poisson bracket as the Cattaneo Poisson bracket \cite{pkg} because the resulting model simplifies to the usual Cattaneo equations when we consider only one-dimensional processes in the linear approximation, see Appendix \ref{sec.Cattaneo.lin}.
The reversible part of the evolution equations, stemming from this bracket, is obtained by formula \eqref{eq.xrev},
\begin{subequations}\label{eq.Cat.rev}
\begin{align}
\label{eq.Cat.s}(\partial_t s)_\rev &= \{s,E\}^{(Cattaneo)} =  -\partial_i E_{w_i}\\
(\partial_t w_i)_\rev &= \{w_i,E\}^{(Cattaneo)} = -\partial_i E_{s} + \frac{1}{s}(\partial_i w_j - \partial_j w_i)E_{w_j}.
\end{align}
The latter equation can be also rewritten as 
\begin{equation}\label{eq.Cat.omega}
(\partial_t \ww)_\rev = -\nabla E_{s} + \frac{1}{s} E_{\ww} \times \oomega,
\end{equation}
where $\oomega = \nabla \times \ww$ is the vorticity of the field $\ww$. From the evolution equation for entropy, we can see that the entropy flux is given by $E_\ww$, and the field $\ww$ can be thus called the conjugate entropy flux.
\end{subequations}

What is the heat flux implied by Equations \eqref{eq.Cat.omega}? If we define the heat flux as the flux of energy, we have to first find the evolution equation for the total energy density $e(s,\ww)$. And since we know evolution equations for both $s$ and $\ww$, we obtain the balance of total energy by chain rule, 
\begin{equation}
\partial_t e = E_s \partial_t s + E_{w_i} \partial_t w_i = -\partial_i \left(E_s E_{w_i}\right),
\end{equation}
which means that the heat flux (or flux of energy) is 
\begin{equation}\label{eq.q}
\qq = E_s E_\ww = -T^2 S_\ww.
\end{equation}
In other words, the heat flux is equal to the product of temperature ($T = E_s$) and entropy flux ($E_\ww$), which is a consequence of the model, not an assumption. Another (equivalent) relation between the heat flux and momentum density is derived later (Equation \eqref{eq.qm}). 

What is the energy $E(s,\ww)$, that is necessary to write the evolution equations in a closed form? It can again be derived by maximization of the entropy for the distribution function \eqref{eq.Sf} subject to the constraints given by the knowledge of the hydrodynamic fields (Equations \eqref{eq.us}). Appendix \ref{sec.maxent} contains details of the calculation, resulting in entropy density 
\begin{equation}\label{eq.sH}
s^H = \frac{4}{3}\left(\frac{6\sigma}{ec}\right)^{1/4} e \frac{3^{3/4}}{2^{7/4}} (3-\chi)^{1/2} (1-\chi)^{1/4},
\end{equation}
where $\sigma = \frac{2 \pi^5 k_B^4}{15h^3c^2}$ is the Stefan-Boltzmann constant and $\chi = \frac{5}{3} - \frac{4}{3} \sqrt{1- \frac{3}{4}\frac{c^2 \mm^2}{e^2}}$ is the Eddington factor \cite{anile}. This entropy is the same as in \cite{mr,larecki}. In the following Section, the reversible evolution is equipped with an irreversible counterpart.


\subsection{Zero-th order irreversible evolution}
By integration of the Callaway model (Equations \eqref{eq.Callaway}) with respect to momentum, we obtain the following zero-th approximation of the irreversible evolution of the $s$ and $\mm$ fields, 
\begin{subequations}
    \begin{align}
        (\partial_t \mm)_{irr} &= -\frac{1}{\tau_R}\mm\\
        (\partial_t s)_{irr} &= \frac{1}{\tau_R E_s}\mm\cdot E_{\mm},
    \end{align}
where the latter equation is obtained from the former by the requirement of energy conservation, $(\partial_t e)_{irr}=0$.
\end{subequations}
These equations can be transformed into the $s,\ww$ variables as follows,
\begin{subequations}
    \begin{align}
        (\partial_t s)_{irr} &= \frac{1}{\tau_R} \frac{\ww\cdot E_{\ww}}{E_s - \frac{1}{s}\ww\cdot E_\ww},\\
        (\partial_t \ww)_{irr} &= -\frac{1}{\tau_R}\ww - \frac{\ww^2}{\tau_R s} \frac{E_\ww}{E_s-\frac{1}{s}E_\ww\cdot\ww}.
    \end{align}
\end{subequations}

The final evolution equations for the hydrodynamic fields $s$ and $\ww$ are the sum of the reversible (Hamiltonian) evolution and irreversible evolution, 
\begin{subequations}\label{eq.Cat.final}
\begin{align}
\partial_t s &= -\partial_i E_{w_i} +\frac{1}{\tau_R} \frac{\ww\cdot E_{\ww}}{E_s - \frac{1}{s}\ww\cdot E_\ww},\\
\label{eq.Cat.final.w}\partial_t w_i &= -\partial_i E_{s} + \frac{1}{s}(\partial_i w_j - \partial_j w_i)E_{w_j} -\frac{1}{\tau_R}\ww - \frac{\ww^2}{\tau_R s} \frac{E_\ww}{E_s-\frac{1}{s}E_\ww\cdot\ww},
\end{align}
which is a nonlinear generalization of the Maxwell-Cattaneo-Vernotte (MCV) equation, see Appendix \ref{sec.Cattaneo.lin}. 
\end{subequations}
Equations \eqref{eq.Cat.final} are a system of quasilinear first-order partial differential equations, and they are also symmetric hyperbolic because they can be rewritten as a system of conservation laws in the $s,\mm$ variables, which is suitable for the Godunov-Boillat theorem \cite{god,boillat,ruggeri-strumia,lageul}. The dissipative part (terms that contain $\tau_R$) can be seen as the zero-th order approximation in the Chapman-Enskog expansion of Equation \eqref{eq.f}. In the following Section, we go for the first Chapman-Enskog approximation, which brings viscous-like terms to the evolution equations.

\subsection{First-order irreversible evolution}\label{sec.ch1}
The first-order Chapman-Enskog approximation of the Boltzmann equation for rarefied gases leads to the Navier-Stokes equations, which contain the viscous terms, on top of the hyperbolic Euler equations \cite{dgm,chapman}. Similarly, the BGK equation, which is actually a form of the here used Callaway model, gives the Navier-Stokes equation as the first Chapman-Enskog approximation \cite{bgk,saint-raymond}. Our goal is now to obtain the viscous-like terms that appear in the momentum equation in the first Chapman-Enskog approximation of the Callaway model also in the phonon hydrodynamics, similarly as in \cite{guo-bgk} but in our state variables.

The Chapman-Enskog expansion starts with the projection of the equation for $f$ (Equation \eqref{eq.f}) to the momentum density $\mm$, 
\begin{subequations}
\begin{equation}\label{eq.m.Q}
\partial_t m_i = -\frac{\partial}{\partial r^j}\left(\int d\pp p_i \frac{c p_j}{|\pp|} f\right)-\frac{1}{\tau_R}m_i.
\end{equation}
For simplicity, we now disregard the upper/lower positions of the indexes, which means that the metric tensor is implicitly present in the calculations.
In order to provide an approximation for the unknown integral on the right hand side, $Q_{ij} = \int d\pp p_i p_j/|\pp| f$, the evolution of that integral is obtained again from the evolution equation for $f$,
\begin{equation}
\partial_t Q_{ij} = -\frac{\partial}{\partial r^k}\left(\int d\pp c \frac{p_i p_j p_k}{|\pp|} f\right) + \frac{1}{\tau_N}\int d\pp \frac{p_i p_j}{|\pp|} (f_N-f) + \frac{1}{\tau_R}\int d\pp \frac{p_i p_j}{|\pp|} (f_R -f).
\end{equation}
\end{subequations}
Appendix \ref{sec.chapman} contains approximate values of the integrals on the right hand side of the latter equation, with which the evolution equation for $Q_{ij}$ becomes
\begin{equation}\label{eq.Q}
\partial_t Q_{ij} \approx 
\frac{\partial}{\partial r^k}\left(\frac{16\pi^5 k_B^4}{75 h^3 (e^*)^5 c^4} (m^*_i\delta^k_j +(m^*)^k\delta_{ji}+m^*_j\delta_i^k)  \right)
-\left(\frac{1}{\tau_R}+\frac{1}{\tau_N}\right)\left(Q_{ij}-\frac{4\pi^5 k_B^4}{15 h^3 (e^*)^4 c^4}\delta_{ij}\right).
\end{equation}

In order to solve approximately \eqref{eq.Q}, we first cast the equation into a dimensionless form. The underlying evolution equation for the distribution function \eqref{eq.f} has the dimensionless form
\begin{equation}
\frac{\partial \hat{f}}{\partial \hat t} = -\mathrm{St} \frac{\hat{\pp}}{|\hat{\pp}|}\cdot \hat{\nabla}\hat{f} + \frac{1}{\mathrm{De}_N} (\hat{f}_{N} - \hat{f})+ \frac{1}{\mathrm{De}_R} (\hat{f}_{R} - \hat{f}),
\end{equation}
where the quantities with hats are dimensionless, 
$\mathrm{St} = \frac{c \bar{t}}{\bar{r}}$ is the Strouhal number, $\bar{t}$ is a typical time value, $\bar{r}$ is a typical spatial dimension, and $\mathrm{De}_N=\tau_N/\bar{t}$ and $\mathrm{De}_R=\tau_R/\bar{t}$ are the Deborah numbers corresponding to the two relaxation times. Because the phonon gas first reaches the quasi-equilibrium manifold (defined by the momentum density and energy density) before it reaches the equilibrium (defined only by the energy), the normal relaxation time is assumed to be much smaller than the resistive one, $\tau_N \ll \tau_R$, which means that $\mathrm{De}_N \ll \mathrm{De}_R$. Therefore, the second moment $Q_{ij}$ can be expanded in powers of the normal Deborah number, 
\begin{equation}
Q_{ij} =  Q^{(0)}_{ij} + \mathrm{De}_N Q^{(1)}_{ij} + \mathcal{O}(\mathrm{De}_N^2).
\end{equation}
Plugging this expansion into the equation for $Q_{ij}$ (Equation \eqref{eq.Q}), the $\mathrm{De}_N^{-1}$ order gives 
\begin{subequations}
\begin{equation}
\mathbf{Q}^{(0)} = \frac{4\pi^5 k_B^4}{15 h^3 (e^*)^4 c^4},
\end{equation}
and the order $\mathrm{De}_N^{0}$ gives 
\begin{equation}
\partial_t Q^{(0)}_{ij} = \partial_k\left(\frac{16\pi^5 k_B^4}{75 h^3 (e^*)^5 c^4} (m^*_i\delta^k_j +(m^*)^k\delta_{ji}+m^*_j\delta_i^k)  \right) - {\bar{t}}^{-1}\left(1+\frac{\mathrm{De}^N}{\mathrm{De}^R}\right)Q^{(1)}_{ij}.
\end{equation}
\end{subequations}
These two orders then combine to 
\begin{equation}\label{eq.Q.approx}
Q_{ij} = \frac{4\pi^5 k_b^4}{15 h^3 (e^*)^4 c^4} + \tau \partial_k\left(\frac{16\pi^5 k_B^4}{75 h^3 (e^*)^5 c^4} (m^*_i\delta^k_j +(m^*)^k\delta_{ji}+m^*_j\delta_i^k)  \right) - \tau \frac{4\pi^5 k_b^4}{15 h^3  c^4}\partial_t  (e^*)^{-4}\delta_{ij}
\end{equation}
where $\tau^{-1} = \tau_R^{-1}+\tau_N^{-1}$ is the effective relaxation time.

The approximate $\mathbf{Q}$ from Equation \eqref{eq.Q.approx} can be now plugged into the momentum equation \eqref{eq.m.Q}. However, since we have already obtained the reversible part of the latter equation precisely, using the Poisson reduction, we use the approximate $\mathbf{Q}$ only in the irreversible part of the equation, 
\begin{equation}\label{eq.me.irr}
(\partial_t m_i)_{\mathrm{irr}} = -\frac{1}{\tau_R}m_i -\partial_j \left(\tau \partial_k\left(\frac{16\pi^5 k_B^4}{75 h^3 (e^*)^5 c^3} (m^*_i\delta^k_j +(m^*)^k\delta_{ji}+m^*_j\delta_i^k) \right) - \tau \frac{4\pi^5 k_b^4}{15 h^3  c^3}\partial_t  (e^*)^{-4}\delta_{ij}\right).
\end{equation}
\begin{subequations}\label{eq.me.full}
The reversible part of the momentum equation is contained in Equation \eqref{eq.m.rev}, and the overall evolution of the momentum density is the sum of th reversible and irreversible parts, 
\begin{align}\label{eq.m.ch1}
\partial_t m_i=&
    \partial_j \left(m_{i} \frac{m^{*j}}{e^*}\right) - s(\mm,e) \partial_i \left(\frac{1}{e^*}\right) + m_{j}\partial_i \frac{m^{*j}}{e^*}\\
 &-\frac{1}{\tau_R}m_i -\partial_j \left(\tau \partial_k\left(\frac{16\pi^5 k_B^4}{75 h^3 (e^*)^5 c^3} (m^*_i\delta^k_j +(m^*)^k\delta_{ji}+m^*_j\delta_i^k) \right) - \tau \frac{4\pi^5 k_b^4}{15 h^3  c^3}\partial_t  (e^*)^{-4}\delta_{ij}\right),\nonumber
\end{align}
where derivatives of energy were replaced with the corresponding derivatives of entropy, $E_\mm = - S_\mm/S_e$, $E_s = (S_e)^{-1}$.
    The right hand side of the momentum equation contains the conjugate variables $\mm^* = \frac{\partial S^{H}}{\partial \mm}$ and $e^* = \frac{\partial S^{H}}{\partial e}$, but we know the entropy (Equation \eqref{eq.sH}), which makes the right hand side explicit in $e$ and $\mm$.

Finally, the momentum equation needs to be supplemented with the energy equation, 
\begin{equation}\label{eq.em}
\partial_t e = \int d\pp c|\pp|\partial_t f = -\partial_i (c^2 m_i),
\end{equation}
\end{subequations}
which also shows that the heat flux is proportional to the momentum density, 
\begin{equation}\label{eq.qm}
\qq = c^2 \mm.
\end{equation}
Equations \eqref{eq.me.full} govern evolution of momentum density and energy density of phonons in the first Chapman-Enskog approximation.  

In order to better see the viscous-like character of the irreversible terms in the momentum equation, let us use the approximate values of the Lagrange conjugate variables \eqref{eq.Lag.approx}, which give that $\mm^*/e^* \approx -\frac{3c^2}{4e}\mm$ and $\frac{16 \pi^5 k_B^4}{75 h^3 (e^*)^4c^3} = \frac{8\sigma}{5c (e^*)^4} \approx \frac{4}{15}e$. Moreover, when $\mm^2\approx 0$, it holds that $2\sigma c^{-1} (e^*)^{-4} \approx e/3$, and the momentum equation then simplifies to
\begin{align}\label{eq.m.ch1.approx}
\partial_t m_i =& -\partial_i p -\partial_j\left(\frac{3c^2}{4e}m_i m_j\right)\\
    &-\frac{1}{\tau_R}m_i +\frac{\tau}{5}\partial_j\left(\partial_j m_i + \partial_i m_j -\frac{2}{3}\nabla\cdot\mm \delta_{ij}\right)\nonumber
\end{align}
with pressure
\begin{equation}\label{eq.p}
    p = -e + \frac{4}{3}e(2\sqrt{y}-1) + \frac{3c^2}{2e}\frac{\mm^2}{1+\sqrt{y}} = \frac{e}{3}-\frac{c^2\mm^2}{4e}+\mathcal{O}(\mm^4) \approx  \frac{e}{3}-\frac{c^2\mm^2}{4e}.
\end{equation}
Equations \eqref{eq.em} and \eqref{eq.m.ch1.approx} are a partial linearization of the full equations for $\mm$ and $e$. However, they still represent a more precise version of the Guyer-Krumhansl equation, since they still contain the convective terms in the phonon momentum density. The precision is enhanced due to the Poisson reduction, which makes it possible to obtain the reversible part of the equation without linearizations. The viscous terms in the momentum equation are similar to those found in \cite{guo-bgk}, except for the terms with $\nabla\cdot\mm$ (here, the viscous part of the Cauchy stress tensor is traceless). Equation \eqref{eq.m.ch1.approx} is similar to the evolution equation for heat flux in \cite{cimmelli-prb}, but it still differs by the presence of the convective term. 

Equations \eqref{eq.em} and \eqref{eq.m.ch1.approx} in one dimension can be linearized around a stationary homogeneous state, $e \approx e_0 + \delta e$ and $\mm = \delta \mm$, which leads to a wave equation (neglecting the dissipative terms),
\begin{equation}\label{eq.secondsound}
\frac{\partial^2 e}{\partial t^2} = \frac{c^2}{3} \frac{\partial^2 e}{\partial x^2}.
\end{equation}
Therefore, the speed of the second sound can be related with the Debye velocity through $c_{\text{(2nd)}} = c/\sqrt{3}$, as reported in \cite{guo2015}.

We would like to solve these equations numerically in the future,
for which we shall need boundary conditions. Although boundary conditions are out of scope of the present paper, let us briefly mention at least some results from the literature.
Apart from no heat flux across an insulated wall $\qq\cdot \nn = 0$, one could consider zero heat flux also along the insulating wall (non-slip condition, $\qq = 0$), or a non-zero value (slip condition) analogous to the no-slip condition of velocity for rarefied gases \cite{alvarez2009,sellitto2015}. These conditions are especially relevant in narrow channels \cite{sellitto2011,lebon2012}. Boundary conditions for the heat flux have also been explored within IVT \cite{szucs-symmetry}, and a derivation of boundary conditions from the BGK equation for phonons can be found in \cite{guo-bgk}.

Let us now discuss some features of Equations \eqref{eq.Cat.final} (obtained by the Poisson reduction and the zero-th Chapman-Enskog approximation).

\subsection{Galilean invariance, inertia, and thermal vortices}
Equations \eqref{eq.Cat.final} have several features that are not present in the usual models for hyperbolic heat conduction, Galilean invariance, inertial effects, and effects of thermal vorticity. First, their reversible part (the Hamiltonian part) is Galilean invariant, as shown in Appendix \ref{sec.Galileo}. Galilean invariance of the irreversible part of the entropy equation is achieved if we count with the possible motion of the crystal (with velocity $\VV$) by taking energy $e=e(s,\ww)+s\VV\cdot\ww$.

Second, the evolution equation for $\ww$ contains inertial terms, which are proportional to $E_\ww$. In other words, if we abruptly change the temperature gradient, it takes some time to change the direction of $\ww$. 

\begin{figure}
\begin{center}
\includegraphics[scale=0.5]{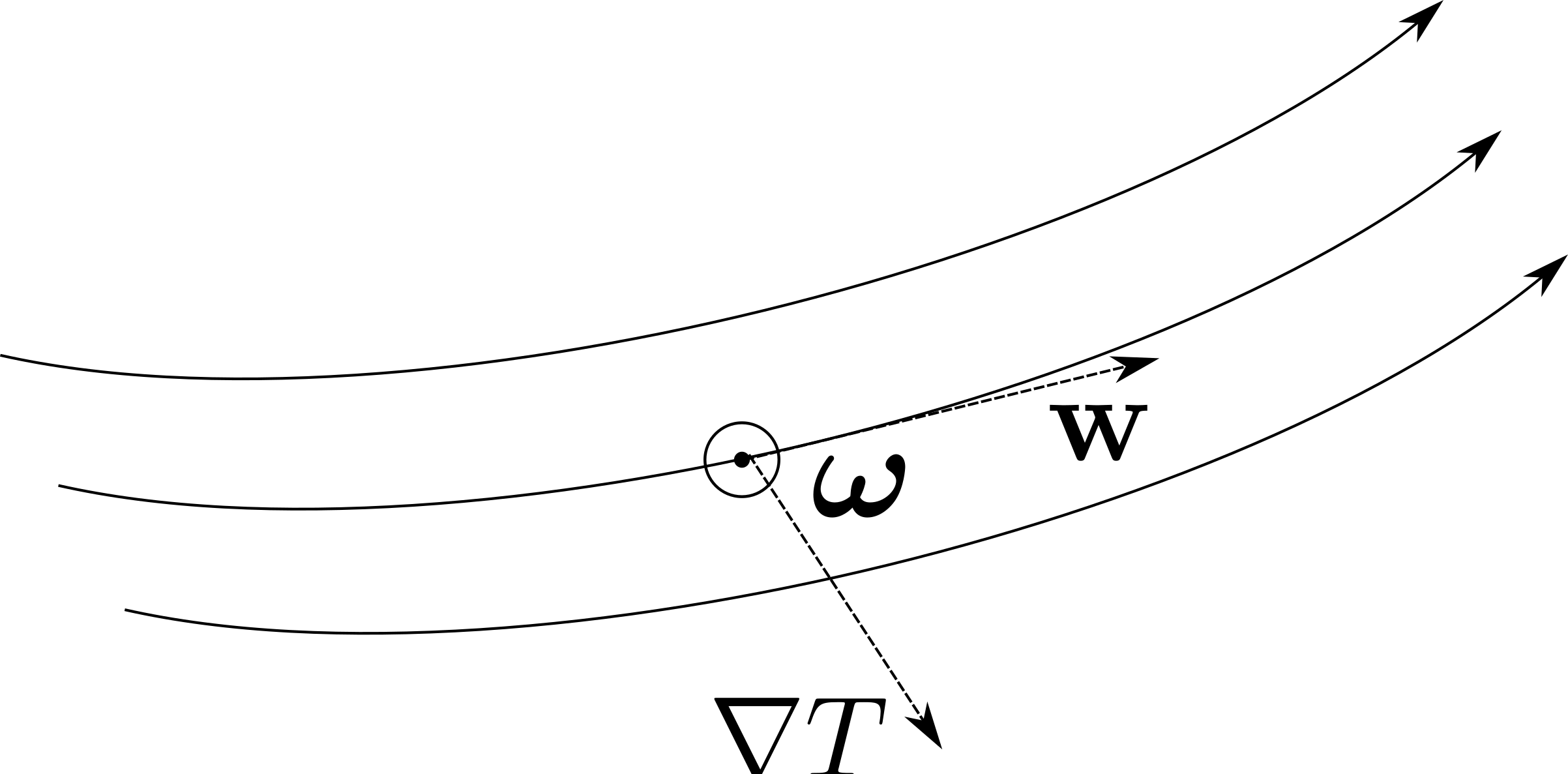}
\caption{\label{fig.crocco} The Fourier-Crocco equation for in the presence of thermal vorticity $\oomega$. The conjugate entropy flux $\ww$, which is proportional to the heat flux, is not parallel with the gradient of temperature. }
\end{center}
\end{figure}

The zero-th Chapman-Enskog approximate evolution of $\ww$ \eqref{eq.Cat.final} also expresses how the thermal vorticity $\oomega=\nabla\times \ww$ affects the evolution of $\ww$ itself. In the stationary state, the evolution equation for $\ww$ yields an analogy of the Crocco theorem \cite{batchelor}
\begin{equation}\label{eq.crocco}
\nabla T = \frac{1}{s}E_{\ww} \times \oomega - \frac{1}{\tau_R}\ww,
\end{equation}
here called the Fourier-Crocco equation.\footnote{Here, we neglect the second irreversible term in the equation for $\ww$ in \eqref{eq.Cat.final} because it is of higher order in $\ww$.}
Figure \ref{fig.crocco} shows this equation in the case of two dimensions. In particular, for the limit of weak dissipation, $\tau_R\rightarrow \infty$, the heat flux is perpendicular to the temperature gradient. On the other hand, in the limit of strong dissipation, $\tau_R\rightarrow 0$, the heat flux aligns with the temperature gradient, as expected from the Fourier theory of heat conduction.

How can be thermal vorticity $\oomega$ created? Consider for instance a body with a gradient of thermal conductivity in the $x$-direction. Thermal conductivity is proportional to $\tau_R^{-1}$ (see the following Section). When we cast the curl operator on the evolution equation for $\ww$ \eqref{eq.Cat.final.w}, we obtain that
\begin{equation}
\partial_t \oomega|_{\oomega = 0} = -\nabla\left(\frac{1}{\tau_R}\right)\times \ww,
\end{equation}
even in the case where $\oomega=0$. Note that now we have to allow for $\tau_R$ not to be a constant.
This means that thermal vorticity $\oomega$ is created when the gradient of $\tau^{-1}_R$ is not aligned with $\ww$. For instance, $\ww$ is created in the $y$-direction (by applying a temperature gradient to the body), some amount of thermal vorticity is created. We expect this behavior in thermal diodes \cite{liliana-diode,wong2021}. The thermal vorticity can be also created at the walls (by boundary conditions) and at sharp corners as in fluid mechanics \cite{batchelor}.

\begin{figure}
     \centering
     \begin{subfigure}{0.5\textwidth}
         \centering
         \includegraphics[width=0.7\textwidth]{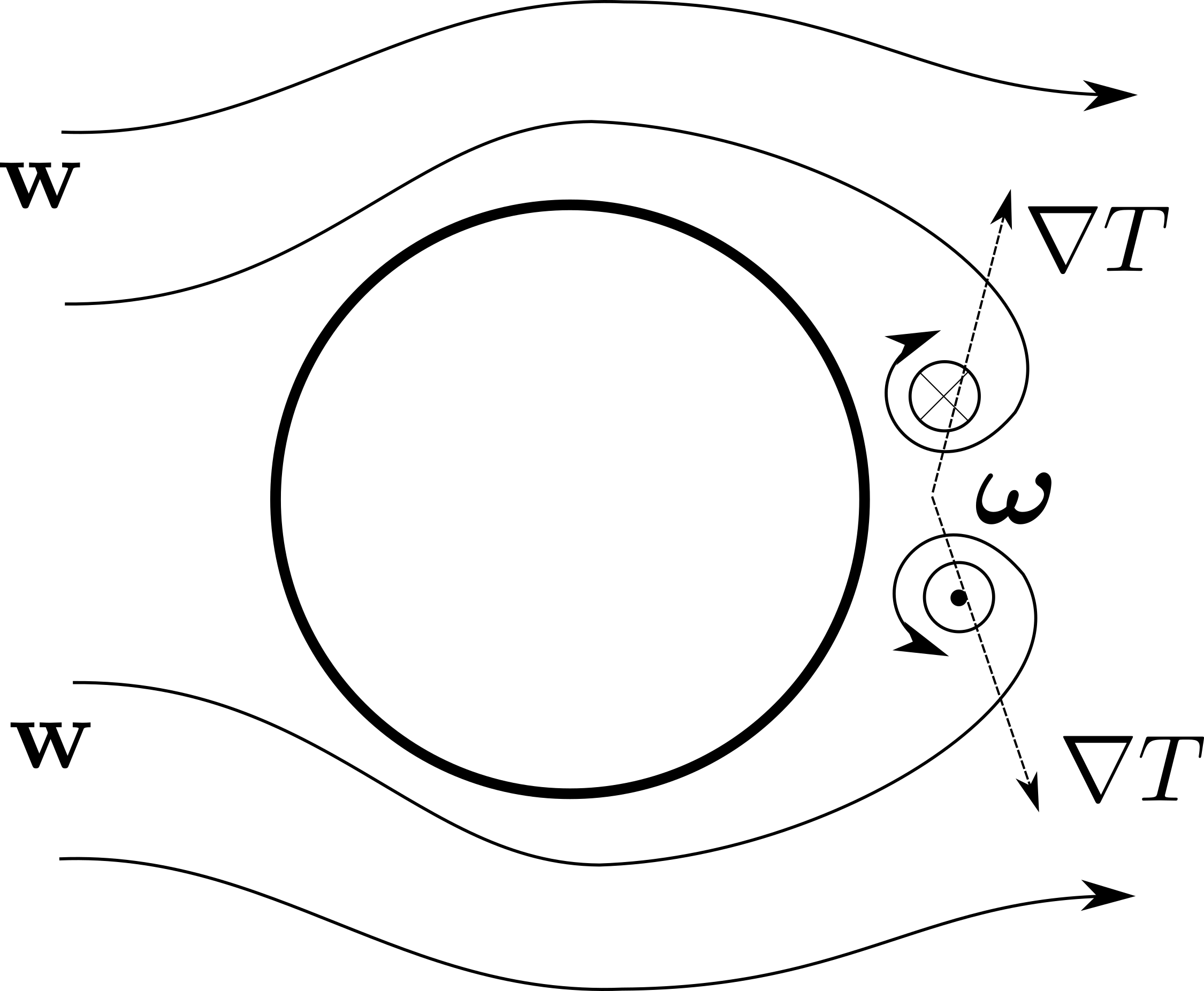}
         \caption{Flow of phonons past an insulated cylinder create a colder spot behind the cylinder.}
         \label{fig.cylinder}
     \end{subfigure}
     \begin{subfigure}{0.5\textwidth}
         \centering
         \includegraphics[width=0.7\textwidth]{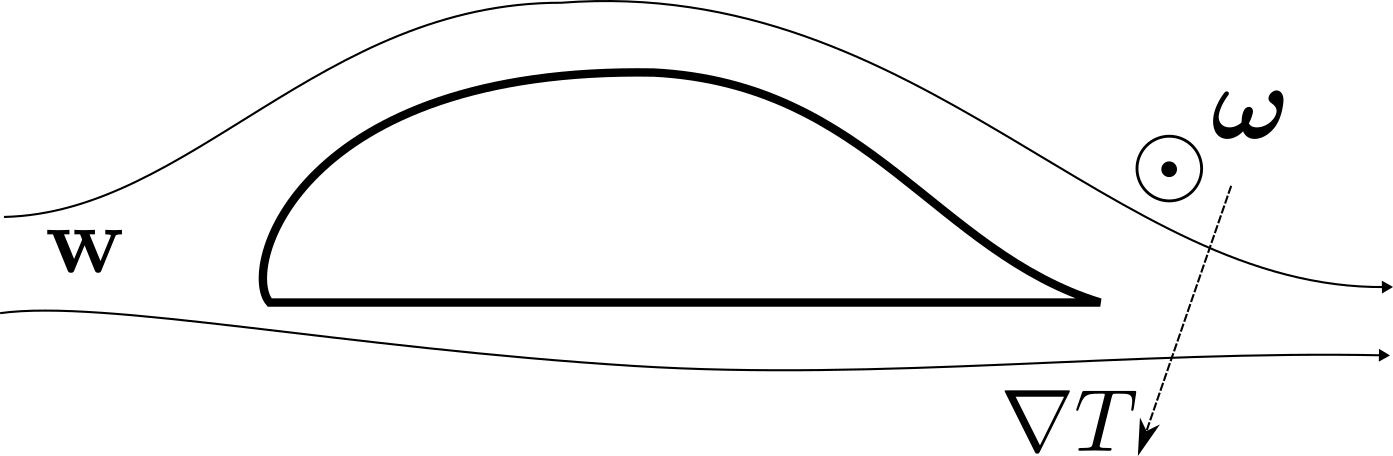}
         \caption{Flow past an airfoil-like insulated body creates a temperature gradient (hotter below).}
         \label{fig.aero}
     \end{subfigure}
\end{figure}

What effects can be expected in the presence of thermal vorticity? Since the Fourier-Crocco equation \eqref{eq.crocco} is similar to the classical Crocco equation from fluid mechanics, we can find inspiration in aerodynamics. Figure \ref{fig.cylinder} shows the flow of phonons around an insulated cylinder. Since the cylinder is insulated, there is no heat flux across its boundary, so that $\ww\cdot\nn=0$ at the boundary. In the stationary state, when the Fourier-Crocco equation applies, we have, moreover, that $\nabla\cdot \mm = 0$, which brings us close to dynamics of incompressible fluids. From the Fourier-Crocco equation, we can expect that the temperature behind the cylinder is lower than the temperature in front of the cylinder. The analogy in classical fluid mechanics would be that the hydrodynamic drag force (or lower pressure behind the cylinder). In other words, an insulated obstacle in a flow of phonons should create a region of lower temperature behind the obstacle. This is, however, still to be verified experimentally.

Similarly, Figure \ref{fig.aero} shows that flow of phonons along an insulated airfoil should create a region or lower temperature above the airfoil and a region of higher temperature below the airfoil. No such effects could be obtained without the non-linear vorticity-dependent terms, that are usually neglected in standard theories of hyperbolic heat conduction \cite{guyer-krumhansl}.

\subsection{Reduction to Fourier heat transport}\label{sec.F}
When we asymptotically expand the conjugate entropy flux $\ww = \ww^{(0)} + \mathrm{De}_R \ww^{(1)} + \mathcal{O}(\mathrm{De}_R^2)$,
then from the evolution equation for $\ww$ we obtain that $\ww^{(0)} = 0$, $\ww^{(1)}=- \nabla T$, and $\ww \approx -\tau_R\nabla T$. The evolution equation for entropy then becomes
\begin{equation}
\partial_t s = \nabla\cdot\left(\tau_R\kappa \nabla T\right) + \frac{\tau_R\kappa}{T} (\nabla T)^2,
\end{equation}
where $\kappa=\frac{3c^2 s^2}{4e}$ is the coefficient in approximated energy $e \approx e^{(0)}(s) + \frac{1}{2}\kappa\ww^2$.
In terms of energy density, this equation reads
\begin{equation}
\partial_t e = -\nabla\cdot(-\tau_R\kappa T \nabla T),
\end{equation}
which means that the heat flux, or flux of energy, is $\qq = -\tau_R T \kappa \nabla T$, that represents Fourier law with thermal conductivity $\lambda=\tau_R T\kappa$. Fourier heat conduction can be thus seen as the first-order approximation of the asymptotic expansion of Equations \eqref{eq.Cat.final} in the limit of strong dissipation.

\subsection{A hypothetical dependence of energy on vorticity}\label{sec.vorticity}
In this Section, we add to the energy a hypothetical dependence on the thermal vorticity $\oomega$, and replace the energy density derived by maximization of the entropy of bosons by 
\begin{equation}
\bar{E} = \int d\rr \bar{e}(s,\ww,\oomega),
\end{equation}
where is $\bar{e}(s,\ww,\oomega)$ is a function of entropy, the conjugate entropy flux, and thermal vorticity. In case of $\oomega=0$, $\bar{e}$ should turn to the standard formula \eqref{eq.sH}, but we leave it undetermined because it is not clear whether such (hypothetical) dependence is possible from the physical point of view. Phonons express collective vibrations of atoms of a lattice, which means that $\oomega$ encodes a sort of angular momentum of the vibrations. A similar energy density can be seen for instance in superfluids, where dependence of energy on vorticity of the superfluid velocity is due to the presence of topological defects (quantum vortices) \cite{Holm-HVBK,Mongiovi2018,sciver,khalatnikov}. We are not aware, however, of any experimental observation of similar vortices in the case of phonons.

Anyway, let us close the paper with exposition of the consequences of the hypothetical dependence of energy density on thermal vorticity. The functional derivative of energy with respect to the $\ww$ field turns to 
\begin{equation}
\frac{\delta \bar{E}}{\delta \ww} = \frac{\partial e}{\partial \ww} + \nabla\times \frac{\partial \bar{e}}{\partial \oomega}.
\end{equation}
The evolution equation for $\ww$ (Equation \eqref{eq.Cat.final.w}) and the equation for curl of $\ww$ become
\begin{subequations}
\begin{align}
\partial_t \ww &= -\nabla T + \frac{1}{s}\frac{\partial \bar{e}}{\partial \ww}\times \oomega + \frac{1}{s}\left(\nabla \frac{\partial \bar{e}}{\partial \ww}\cdot\oomega - \omega\cdot \nabla \frac{\partial \bar{e}}{\partial \ww}\right) - \frac{1}{\tau_R}\ww\\
\partial_t \oomega &= \nabla\cdot\left(\frac{1}{s}\left(\frac{\partial \bar{e}}{\partial \ww}\otimes\oomega - \oomega \otimes \frac{\partial \bar{e}}{\partial \ww}\right)\right) + \nabla\times\left(\frac{1}{s}\left(\nabla \frac{\partial \bar{e}}{\partial \ww}\cdot\oomega - \omega\cdot \nabla \frac{\partial \bar{e}}{\partial \ww}\right)\right) + \ww\times \nabla\left(\frac{1}{\tau_R}\right) - \frac{1}{\tau_R}\oomega.
\end{align}
\end{subequations}
The equation for thermal vorticity has thus two fluxes (terms under the $\nabla$ operator), a source term (with gradient of $\tau_R$), and a sink term (proportional to $\oomega$). In particular, if $\oomega$ is zero initially and if there is no gradient in $\tau^{-1}_R$, then $\oomega$ remains zero. On the other hand, if there is a gradient in $\tau_R$, as in case of a thermal diode \cite{liliana-diode,wong2021}, then some amount of $\oomega$ is being created until a balance with the sink term is reached.

Heat flow also exhibits hydrodynamics behaviour in superfluid helium-4; for instance, in the laminar regime in counterflow along a cylindrical tube or along tubes or rectangular or square cross section, the behaviour of the heat flow is analogous to that of viscous fluids \cite{Mongiovi1991,Mongiovi2017,Mongiovi2018}. In the two-fluid model \cite{Tisza,landau-helium,khalatnikov,Holm-HVBK} this may be easily interpreted, as the heat flow is transported by the viscous component of the fluid \cite{pof2021}. Exploration of a possible von Kármán vortex street when an insulating cylinder perpendicular to the heat flux is introduced in a parallelepiped channel could be of interest. However, in superfluid helium the circulation of the vortices is quantized (due to the requirement that the $\nabla\times\ww=0$ almost everywhere), which makes it different from the phonon hydrodynamics explored in the present paper.

\section{Conclusion}
In this paper, we begin with the evolution equation of distribution functions of phonons with two transversal and one longitudinal polarizations (actually kinetic theory of mixtures for $f_{T+}$, $f_{T-}$, and $f_L$). The energy is simplified to the energy of acoustic phonons in the first Brillouin zone, and the entropy is obtained as the entropy of an ideal bosonic gas in kinetic theory. 

This description is then reduced to hydrodynamics with momentum densities and entropy densities of the particular polarizations ($\mm_\alpha$ and $s_\alpha$ for $\alpha \in \{T+,T-,L\}$). The underlying Poisson bracket is obtained by reduction from the kinetic Poisson bracket and the entropy is derived from the entropy in kinetic theory by means of MaxEnt. The kinetic theory of polarized phonons is also reduced to the overall kinetic theory of phonons with distribution function $f=\sum_\alpha f_\alpha$ (disregarding polarizations), where the Poisson bracket is again obtained by reduction from the kinetic Poisson bracket, and entropy is obtained again by MaxEnt. 

Finally, these levels of description are further reduced to phonon hydrodynamics described by state variables $\mm=\sum_\alpha \mm_\alpha$ and $s=\sum_\alpha s_\alpha$ (momentum density and entropy density). The Poisson bracket and entropy are again derived by reduction from their more detailed counterparts. Apart from the zero-th Chapman-Enskog approximation, which leads to hyperbolic equations, we also carry out the first Chapman-Enskog approximation, which adds higher-order (Laplacian) terms to the evolution equation for phonon momentum density. It turns out more practical to reformulate the hydrodynamic equations on this level of description to variables $\ww=\mm/s$ and $s$, since then we can better see the relation with standard theories of heat transport. The obtained evolution equations are Galilean invariant, unlike usual theories of heat transport, and they depend on the thermal vorticity, $\oomega = \nabla\times \ww$. 

In particular, even in the stationary case, the heat flux turns out to be not aligned with the gradient of temperature, becoming proportional to it only in the limit of strong dissipation. In the one-dimensional case and after linearization, the theory simplifies to the Maxwell-Cattaneo-Vernotte equation. The thermal vorticity is created if gradients in thermal conductivity are present that are not parallel with the heat flux. In the presence of thermal vorticity, we can expect various phenomena similar to aerodynamics, like a Fourier-Crocco equation \eqref{eq.crocco}, drop of temperature (instead of pressure) behind and obstacle, or temperature gradient perpendicular to the flow around an airfoil (instead of aerodynamic lift).

In future, we would like to investigate the resulting equations numerically in order to quantitatively observe the here proposed phenomena (for instance the aerodynamic analogies). We would also like to explore instabilities of the equations.
It would also be interesting to apply the present formalism to the examples illustrated by \cite{Zhang2021}. 

\section*{Acknowledgment}
MP is grateful to Alberto Montefusco for asking the questions how exactly the dynamics of the $s$ and $\ww$ fields correspond to the classical Cattaneo equation. MP is also grateful to Miroslav Grmela for the countless discussions of the GENERIC framework, to Václav Klika for discussing the Onsager-Casimir reciprocal relations, and to Ilya Peshkov for discussing hyperbolic PDE's.
MP was supported by project No. UNCE/SCI/023 of Charles University Research program and by the Czech Science Foundation (project no. 20-22092S). 

\appendix

\section{Maximum entropy principle for energy and entropy}\label{sec.maxent}
Entropy of phonons, which are bosons, is expressed in terms of their distribution function \cite{landau5}
\begin{equation}\label{eq.sf1}
S^{\text{kinetic}}(f) = -\frac{k_B}{h^3}\int d\rr \int d\pp \left(h^3 f \ln(h^3 f)-(1+h^3f)\ln(1+h^3f)\right),
\end{equation}
where $h$ is the Planck constant. The goal is now to maximize the entropy with the constraints given by the knowledge the energy density and momentum density fields. This procedure is equivalent to that in \cite{mr}, with the difference that here we obtain also the entropy for an intermediate level of description where the various polarizations of phonons have their own hydrodynamic entropies.

Indeed, because there are two transverse and one longitudinal polarizations of phonons, we actually have three distribution functions $f_{T+}$, $f_{T-}$, and $f_L$, to which correspond three entropies, given by formula \eqref{eq.sf1} for each of the distribution functions. The total entropy is then the sum of the entropies for each polarization, Equation \eqref{eq.Sfa}. In order to obtain analytical results, we neglect phonon-phonon interactions, as well as phonon-crystal interactions. We shall be constrained to the first Brillouin zone, where all the polarizations have the same formula for energy. Moreover, we stay restricted only to the case of acoustic phonons. This will give us the hydrodynamic entropy of the mixture of the three types of acoustic phonons. Eventually, we shall apply MaxEnt once more to obtain the overall hydrodynamic entropy of all acoustic phonons, which turns out to be the same as is \cite{mr}.

The projection to the fields of momentum density and energy density is
\begin{subequations}\label{eq.f.me}
\begin{align}
\mm(\rr) &= \int d\pp\, \pp f\\
e(\rr) &= \int d\pp c|\pp|f,
\end{align}
\end{subequations}
where $c$ is the speed of sound and $|\pp|$ is the Euclidean norm of the momentum, $\pp=\hbar \mathbf{k}$.
By maximization of entropy \eqref{eq.Sfa} subject to the constraints given by fields $e$ and $\mm$, 
\begin{equation}
\frac{\delta}{\delta f}\left(-S^{\text{kinetic}} + \int d\rr \mm^*(\rr)\cdot\int d\pp \pp f + \int d\rr e^*(\rr) \int d\pp c|\pp|f\right)=0,
\end{equation}
we obtain the MaxEnt estimate of the distribution function 
\begin{equation}\label{eq.f.maxent}
f_{\text{MaxEnt}}(\mm,e) = \frac{1}{h^3}\frac{1}{e^{\mm^*\cdot p/k_B}e^{e^*c|\pp|/k_B}-1},
\end{equation}
where $\mm^*$ and $e^*$ are the corresponding Lagrange multipliers. The hydrodynamic formula for entropy of phonons is then given by $S^{\text{kinetic}}(f_{\text{MaxEnt}})$, but in order to express it in terms of state variables $\mm$ and $e$, we have to first express the Lagrange multipliers in terms of the state variables.

By integration of equations \eqref{eq.f.me} with the MaxEnt estimate of the distribution function \eqref{eq.f.maxent}, we obtain that 
\begin{subequations}\label{eq.multipliers}
\begin{align}
m_i &= -\frac{16}{45}\frac{k_B^4 \pi^5}{h^3 (c e^*)^5} \frac{m^*_i}{(1-\gamma^2)^3}\\
e &= \frac{4}{45}\frac{\pi^5 k_B^4}{(e^*)^4 c^3 h^3}\frac{\gamma^2+3}{(1-\gamma^2)^3},
\end{align}
\end{subequations}
where $\gamma = \frac{|\mm^*|}{e^*c}$. Note that is is not trivial to carry out the integration and several tricks are necessary. Apart from standard techniques like swapping integrals and series or transformation to the spherical coordinates, it is useful to rotate the coordinates so that $\mm^*$ becomes $(0,0,|\mm^*|)$.
Equations \eqref{eq.multipliers} have to be solved so that the multipliers, $\mm^*$ and $e^*$, are expressed in terms of the state variables, $\mm$ and $e$. In particular, the equation imply that
\begin{equation}
\frac{\beta^2}{16}(\gamma^4+6\gamma^2+9) = \gamma^2,
\end{equation}
where $\beta = \frac{|\mm|c}{e}$, which gives that 
\begin{equation}
\gamma^2 = 3 \frac{1-\sqrt{y}}{1+\sqrt{y}}
\end{equation}
with $y = 1- \frac{3}{4}\beta^2$. Note that this quadratic equation has two roots, one of which is has no physical meaning because it would lead to a singularity at $\beta=0$ (equilibrium, where $\mm=0$). Since $\beta$ is a function of the state variables, the Lagrange multipliers are now expressed in terms of the state variables, $\mm$ and $e$, as
\begin{subequations}\label{eq.Lagrange}
\begin{align}
(e^*)^4 &= \frac{3\sigma}{2ec}\frac{(1+\sqrt{y})^2}{(2\sqrt{y}-1)^3}\\
\mm^* &= -3\frac{c^3(3\sigma)^{1/4}}{(2ec)^{5/4}}\frac{1}{\sqrt{1+\sqrt{y}}}\frac{1}{(2\sqrt{y}-1)^{3/4}}\mm.
\end{align}
\end{subequations}
Near the equilibrium, where $y=1-\frac{3}{4}\frac{\mm^2 c^2}{e^2} \approx 1$, the multipliers are approximated by 
\begin{equation}\label{eq.Lag.approx}
    e^*\approx \left(\frac{6\sigma}{ec}\right)^{1/4}\qquad\text{ and }\qquad \mm^*\approx -3\frac{c^3(3\sigma)^{1/4}}{\sqrt{2}(2ec)^{5/4}}\mm.
\end{equation}

The hydrodynamic entropy is then obtained by plugging $f_{\text{MaxEnt}}(\mm,e)$ to the kinetic formula for entropy $S^{\text{kinetic}}$, which after some calculations results in 
\begin{equation}
S^{\text{kinetic},\alpha}(f_{\text{MaxEnt}})(\mm_\alpha,e_\alpha) = \int d\rr \frac{4}{3}\left(\frac{2\sigma}{e_\alpha c}\right)^{1/4} e_\alpha \frac{3^{3/4}}{2^{7/4}} (3-\chi_\alpha)^{1/2} (1-\chi_\alpha)^{1/4},
\end{equation}
$\alpha\in\{T+,T-,L\}$, $\sigma = \frac{2 \pi^5 k_B^4}{15h^3c^2}$ is the Stefan-Boltzmann constant, and $\chi_\alpha = \frac{5}{3} - \frac{4}{3} \sqrt{1- \frac{3}{4}\frac{c^2 \mm_\alpha^2}{e_\alpha^2}}$ is the Eddington factor.
Because we actually have three polarizations, $T+$, $T-$, and $L$, we also have three entropies $S^{\text{kinetic},\alpha}$, while the total entropy is their sum, 
\begin{equation}\label{eq.SHP}
S^{\text{HP}} = \sum_{\alpha}S^{\text{kinetic},\alpha}(f_{\text{MaxEnt}})(\mm_\alpha,e_\alpha),
\end{equation}
referred to as the hydrodynamic (H) entropy of polarized (P) phonons. 
In other words, we have obtained the entropy for the level of description with state variables $\mm_{T+}$, $\mm_{T-}$, $\mm_{L}$, $e_{T+}$, $e_{T-}$, and $e_{L}$.

The level of description where the various polarizations of acoustic phonons have their own momentum and energy allows for description of processes like change of polarization of the phonons. However, when such processes are not important, one may opt out for a simpler level of description, where only the overall momentum and energy, 
\begin{subequations}\label{eq.me}
\begin{align}
\mm &= \mm_{T+}+\mm_{T-}+\mm_L\\
e &= e_{T+}+e_{T-}+e_L,
\end{align}
\end{subequations}
are known. What is the entropy on such reduced level of description? The answer is given again by maximization of the total entropy \eqref{eq.SHP} subject to the constraints given by projection \eqref{eq.me}, that is by 
\begin{equation}
\frac{\delta S^{\text{kinetic},\alpha}}{\delta \mm_\alpha}\Big|_{\mm^{\text{MaxEnt}}_\alpha,e^{\text{MaxEnt}}_\alpha} = \mm^* 
\quad\text{and}\quad 
\frac{\delta S^{\text{kinetic},\alpha}}{\delta e_\alpha}\Big|_{\mm^{\text{MaxEnt}}_\alpha,e^{\text{MaxEnt}}_\alpha} = e^*, 
\quad \forall \alpha.
\end{equation}
Since all the formulas for $S^{\text{kinetic},\alpha}$ share the same form, their solutions are the same and the MaxEnt estimates of $\mm_\alpha$ and $e_\alpha$ are the same for all polarizations $\alpha$. Finally, the overall hydrodynamic entropy is obtained by plugging the MaxEnt estimates to the formula for the total entropy \eqref{eq.SHP},
\begin{equation}
S^{\text{H}}(\mm,e) = S^{\text{HP}}(\mm^{\text{MaxEnt}}_\alpha,e^{\text{MaxEnt}}_\alpha) = \int d\rr \frac{4}{3}\left(\frac{6\sigma}{ec}\right)^{1/4} e \frac{3^{3/4}}{2^{7/4}} (3-\chi)^{1/2} (1-\chi)^{1/4},
\end{equation}
where $\chi = \frac{5}{3} - \frac{4}{3} \sqrt{1- \frac{3}{4}\frac{c^2 \mm^2}{e^2}}$ is the overall Eddington factor.

In the equilibrium, we obtain the equilibrium volumetric entropy density 
\begin{equation}\label{eq.S.eq}
s^{\text{E}} = \frac{4}{3}\left(\frac{6\sigma}{c}\right)^{1/4} e^{3/4},
\end{equation}
which means that $e = \frac{6\sigma}{c}T^4$ with temperature $T=\left(\frac{\partial s^{\text{E}}}{\partial e}\right)^{-1}$.

\section{Simplification to the Cattaneo equation}\label{sec.Cattaneo.lin}
How does the dynamics for $s$ and $w$ \eqref{eq.Cat.final} correspond to the classical Cattaneo equation \cite{catt}?
Because Cattaneo equation is a linear equation for temperature in one dimension. Therefore, we shall be restricted to the one-dimensional case, $\ww=(w,0,0)$, we shall transform the $(s,\ww)$ variables to $(T,\ww)$, and finally we will neglect all nonlinear terms.

The derivative of energy $e$ transform to the derivatives of the free energy $f=e- Ts$ as
\begin{equation}
    \left(\frac{\partial e}{\partial s}\right)_\ww = T, 
    \quad
    \left(\frac{\partial e}{\partial \ww}\right)_s = \left(\frac{\partial f}{\partial \ww}\right)_T.
\end{equation}
Derivatives of the free energy (in the $(T,w)$ variables) will be denoted by subscripts.
Using the chain rule, Equations \eqref{eq.Cat.final} become
\begin{subequations}\label{eq.Cat.Tw}
    \begin{align}
        &\left(\frac{\partial s}{\partial T}\right)_\ww \partial_t T +
        \left(\frac{\partial s}{\partial w}\right)_T \cdot \partial_t w 
         = -\partial_x f_w + \frac{1}{\tau_R T} f_w^2\\
        &\partial_t w = -\partial_x T-\frac{1}{\tau_R} f_w.
    \end{align}
\end{subequations}

Let us assume, for simplicity, that we are restricted to a narrow temperature range around a temperature $T_0$ and low values of $w$, so that the free energy is quadratic as well
\begin{equation}
    f(T,w) \approx f_0 -s_0 (T-T_0) - \frac{c_V}{2T_0} (T-T_0)^2+ \frac{1}{2}\kappa w^2,
\end{equation}
where $f_0 = f(T_0,0)$, $s_0 = -f_T$ at $T_0$ and $w=0$, and $c_V = T \frac{\partial s}{\partial T}$ is the heat capacity.
Then we have, in particular, that $f_{TT} = -c_V/T_0 = -\beta$, $\beta$ being a positive constant, $f_{wT}=0$, and $f_{ww} = \kappa$. Note that the non-negative coefficient $\kappa$ is a measure of the non-Fourier behavior. 
Equations \eqref{eq.Cat.Tw} can be then rewritten as
\begin{subequations}
\begin{align}
    \label{eq.Cat.T}\beta \partial_t T &= -\partial_x f_w + \frac{1}{\tau_R T} f_w^2= -\kappa \partial_x w+ \frac{\kappa^2}{\tau_R T} w^2\\
    \partial_t w &= -\partial_x T - \frac{1}{\tau_R} f_w = -\partial_x T -\frac{\kappa}{\tau_R} w.
\end{align}
\end{subequations}

Partial time-derivative of Equation \eqref{eq.Cat.T} then yields, assuming that $\tau_R=const$, 
\begin{align}
    \beta \partial_t \partial_t T &= -\kappa\partial_x \left(-\partial_x T - \frac{\kappa}{\tau_R} w\right) 
    +\frac{\kappa^2}{\tau_R T} \left(-\frac{\partial_t T}{T}w^2 + 2w \partial_t w\right) \nonumber\\
    &= \kappa\partial_x\partial_x T -\frac{\kappa\beta}{\tau_R} \partial_t T + \OBig(\kappa^2\cdot w).
\end{align}
Assume that the terms of order $\kappa^2 w$ (not necessarily the terms or order $\kappa^2 \partial_x w$) are smaller than the remaining terms, we obtain the classical Maxwell-Cattaneo-Vernotte telegrapher's equation,
\begin{equation}
    \beta \partial_t \partial_t T + \frac{\kappa\beta}{\tau_R} \partial_t T = \kappa\partial_x \partial_x T.
\end{equation}
In summary, after the transformation to variables $T$ and $\ww$, we obtain in the one-dimensional case and after linearization the classical Maxwell-Cttaneo-Vernotte equation. However, Equations \eqref{eq.Cat.final} are more general.

\section{Galilean invariance of phonon hydrodynamics}\label{sec.Galileo}
Consider an observer $\mathcal{O}$ in an inertial laboratory frame, who measures time and space positions in $t$ and $\rr$. Consider, moreover, another observer $\mathcal{O}'$ passing by the laboratory with a constant velocity $\VV$. Observer $\mathcal{O}'$ measures time and space positions $t$ and $\rr'=\rr-\VV t$. 

An object with mass $M$ at position $\rr'$ is at position
$\rr = \rr' + \VV t$
in the laboratory frame. By differentiation with respect to time, the velocities become related by
	$\vv = \vv' + \VV$.
Momentum of the object in the laboratory frame, $\pp = M \vv$, relates to the momentum in the $\mathcal{O}'$ frame as
$\pp = \pp' + M \VV$. In particular, if the object has no mass, as a phonon, then its momentum does not change, that is $\mm=\mm'$.
The $\ww$ field is equal to $\mm/s$, and entropy itself is Galilean invariant ($s=s'$), which means that the $\ww$ field does not change under Galilean transformations, $\ww=\ww'$.

Energy of the object transforms as
	$e = \frac{1}{2}M \vv^2 = e' + \pp'\cdot \VV + \frac{1}{2}M \VV^2$,
which is a general formula for Galilean transformations of energy \cite{LBLL}. In particular, if the object has no mass (as a phonon), then the formula simplifies to $e' = e - \mm \cdot\VV$,
where $e$ stands for the volumetric energy density and $\mm$ for the volumetric momentum density of phonons. 
In terms of the $\ww$ field, the energy density transforms as $e'=e-s\ww\cdot\VV$, which means that derivatives of energy transform as
\begin{equation}
\frac{\partial e}{\partial s} = \frac{\partial e'}{\partial s'} + \ww'\cdot\VV
\quad\text{and}\quad
\frac{\partial e}{\partial \ww} = \frac{\partial e'}{\partial \ww'} + s\VV.
\end{equation}

How do the time-and-space derivatives transform? For any function $s(t,\rr)$ it holds that
\begin{subequations}
\begin{align}
	\left(\frac{\partial s}{\partial t}\right)_{\rr} &= \left(\frac{\partial s}{\partial t}\right)_{\rr'} - \left(\frac{\partial s}{\partial r'^j}\right)_t \left(\frac{\partial r'^j}{\partial t}\right)_{\rr'} = 
	 \left(\frac{\partial s}{\partial t}\right)_{\rr'} - V^j \left(\frac{\partial s}{\partial r'^j}\right)_t \\
	\left(\frac{\partial s}{\partial \rr'}\right)_{t} &= \left(\frac{\partial s}{\partial \rr'}\right)_t \left(\frac{\partial \rr'}{\partial \rr}\right)_{t} = 
	 \left(\frac{\partial s}{\partial \rr'}\right)_t.
\end{align}
\end{subequations}
The reversible part of the evolution equations for the fields $s$ and $\ww$ then transform to
\begin{equation}
\left(\frac{\partial s}{\partial t}\right)_{\rr'} \cancel{-V^j \frac{\partial s}{\partial r'^j}}
=
-\frac{\partial}{\partial r'^i}\left(\frac{\partial e'}{\partial w'_i} + V^i s\right) = 
-\frac{\partial}{\partial r'^i}\left(\frac{\partial e'}{\partial w'_i}\right) \cancel{- V^i \frac{\partial s}{\partial r'^i}}
\end{equation}
and 
\begin{equation}
\left(\frac{\partial w'_i}{\partial t}\right)_{\rr'} \cancel{-V^j \frac{\partial w'_i}{\partial r'^j}}
=
-\frac{\partial}{\partial r'^i}\left(\frac{\partial e'}{\partial s'} + \cancel{V^i w'_i}\right) 
+\frac{1}{s'}\left(\frac{\partial w'_j}{\partial r'^i} - \frac{\partial w'_i}{\partial r'^j}\right)\left(\frac{\partial e'}{\partial w'_j} + \cancel{V^j s}\right),
\end{equation}
which means that the reversible part of the equations is Galilean invariant. Note, however, that without the last term on the right hand side, the evolution equation for $\ww$ would not be Galilean invariant. The convective terms that we get by the Poisson reduction thus make the equations Galilean invariant.

\section{Reversibility, irreversibility, and Onsager-Casimir reciprocal relations}\label{sec.app.OCRR}
When talking about reversible and irreversible evolution, we shall first define what we mean by (ir)reversibility. We use the definition based on the time-reversal transformation (TRT) \cite{meixner,casimir1945}. TRT inverts velocities of all particles and the magnetic field, and a quantity is called even with respect to TRT (parity $\Par = 1$) if it is not affected by that inversion. On the other hand, a quantity is odd (parity $\Par = -1$) if its sign flips under TRT. Note that there are also quantities that are neither even nor odd, as for instance distribution functions, but the concept of parity still works in a geometrical sense \cite{pre15}. Most state variables, however, have a definite parity, for instance energy, entropy, and density are even, whereas momentum and velocity are odd. 

Also, parity of a quantity can change. For instance, parity of the heat flux, or the conjugate entropy flux $\ww$, is odd if $\ww$ is among the state variables. On the other hand, after the reduction to the Fourier heat conduction (Section \ref{sec.F}), where only the energy density is the state variable and $\ww$ is proportional to the temperature gradient, parity of $\ww$ becomes even (the same as parity of the temperature gradient). 

Consider now an evolution equation written in such a form that the time-derivative of a state variable is on the left hand side. The right hand side of the evolution equation is then interpreted as a component of a vector field $X^i$. The part of the right hand side that transforms under TRT in the same way as the left hand side is called reversible, while the part that flips the sign is irreversible. In order to generate reversible dynamics, the Poisson bivector must fulfill that
\begin{equation}\label{eq.Par.L}
TRT(L^{ij}) = -\Par(x^i)\Par(x^j)L^{ij}.
\end{equation}
Since energy is even, this property indeed ensures that the Hamiltonian part of the GENERIC evolution transforms as the left hand side of the evolution equations and is thus reversible in the above sense. Property \eqref{eq.Par.L} is fulfilled by the canonical Poisson bracket, from which the reversibility is inherited by the Boltzmann Poisson bracket and all brackets obtained by its projection (brackets within this manuscript) \cite{pre15}. 

Entropy is required to be a Casimir of the Poisson bracket, that is $\{S,A\}=0$ for all $A$, and the Hamiltonian evolution then does not produce entropy (it can only redistribute it). The dissipative part is usually required not to produce or destroy energy, so that $\langle E_{\xx}, \Xi_{\xx^*}\rangle|_{\xx^* = S_\xx}=0$. These degeneracy requirements ensure that energy is conserved by the irreversible part while entropy being conserved by the reversible part. 

The dissipation potential is then typically required to be convex in order to produce entropy, $\dot{S} = \langle S_\xx, \Xi_{\xx^*}\rangle|_{\xx^* = S_\xx}\geq 0$, see e.g. \cite{adv}. Note, however, that convexity is just a sufficient condition of the positivity of entropy production, but not necessary \cite{nonconvex}.

The dissipation potential is supposed to be even with respect to TRT, which means that it generates irreversible evolution. A part of the nonlinear Onsager-Casimir reciprocal relations (OCRR) can be seen as the symmetry of the Hessian of the dissipation potential \cite{gyarmati}. However, to be closer to the historical formulation of OCRR, let us assume, for a moment, that the dissipation potential is quadratic as in Equation \eqref{eq.Xi.quad}. 

The requirement that $\Xi$ be even with respect to TRT means that 
\begin{equation}\label{eq.Par.M}
TRT(M^{ij}) = \Par(x^i)\Par(x^j)M^{ij}
\end{equation}
and the gradient part of the GENERIC evolution then indeed generates irreversible evolution in the above sense. Operator $\MM$ is often referred to as the dissipative matrix, it is symmetric, and it can be obtained as the second functional derivative of the dissipation potential,
\begin{equation}
M^{ij} = \frac{\delta^2 \Xi}{\delta x^*_i \delta x^*_j}.
\end{equation}

We are now in position to recall a generalized version of OCRR \cite{hco,pkg}. GENERIC evolution can be rewritten in the form 
\begin{equation}\label{eq.K}
\partial_t x^i = \underbrace{\left(T_0 L^{ij} - M^{ij}\right)}_{=K^{ij}}\frac{\partial \Phi}{\partial x^i}
\end{equation}
with thermodynamic potential $\Phi = -S + \frac{1}{T_0}E$ ($T_0$ a constant). 
The operator $\KK$ in front of the derivatives of the potential $\Phi$ provides coupling between the state variables and it plays the role of the Onsager-Casimir matrix of phenomenological coefficients. Let us now explicitly discuss particular cases of the possible couplings. Due to properties \eqref{eq.Par.L} and \eqref{eq.Par.M}, we obtain the OCRR \cite{hco}: 
\begin{itemize}
\item Variables with the same parities, $\Par(x^i) = \Par(x^j)$, are coupled through an operator symmetric with respect to the simultaneous TRT and transposition.
\item Variables with opposite parities, $\Par(x^i) = -\Par(x^j)$, are coupled through an operator skew-symmetric with respect to the simultaneous TRT and transposition.
\end{itemize}
These properties are satisfied by both the reversible and irreversible parts of GENERIC, in particular by the Hamiltonian mechanics, as can be checked directly. Since GENERIC can be seen as a consequence of Hamiltonian mechanics \cite{JSP2020}, OCRR can be seen as a consequence of Hamiltonian mechanics as well. 

Let us now illustrate TRT on phonon hydrodynamics. The Poisson bivector for the $(s,\ww)$ variables is
\begin{subequations}
\begin{align}
    L^{s s} &= \{s(\rr'),s(\rr'')\} = 0\\
    L^{s w_k} &= \{s(\rr'),w_k(\rr'')\} = \int d\rr \partial_i \delta(\rr'-\rr)\delta_{ki}\delta(\rr''-\rr) = \partial''_k\delta(\rr'-\rr'')\\
    L^{w_k s} &= \{w_k(\rr'),s(\rr''),\} = -\int d\rr \partial_i \delta(\rr''-\rr)\delta_{ki}\delta(\rr'-\rr) = -\partial'_k\delta(\rr''-\rr')\\
    L^{w_k w_l} &= \{w_k(\rr'),w_l(\rr''),\} = \int d\rr \frac{1}{s(\rr)}(\partial_i w_j-\partial_j w_i)|_{\rr}\delta_{ki}\delta(\rr'-\rr)\delta_{lj}\delta(\rr''-\rr)\nonumber\\
    &= \frac{1}{s(\rr')}(\partial'_k w_l-\partial'_l w_k)|_{\rr'}\delta(\rr''-\rr').
\end{align}
\end{subequations}
We can check that the evolution generated by the bivector is reversible, since it satisfies condition \eqref{eq.Par.L}. Coupling between variables of the same parity, for instance $L^{\ww\ww}$, is symmetric with respect to the simultaneous transposition and time-reversal. Coupling between variables with opposite parities, here $L^{s\ww}$ and $L^{\ww s}$, is skew-symmetric in the same sense. Hamiltonian mechanics thus provides both the symmetric and skew-symmetric coupling.


The irreversible evolution in Equations \eqref{eq.Cat.final} can be considered as gradient dynamics with dissipation potential
\begin{equation}
\Xi = \frac{1}{2}\int d\rr \xi(\ww^*)^2,
\end{equation}
($xi$ being a constant) and the dissipative matrix can be obtained as the second derivative of the dissipation potential,
\begin{equation}
\MM = 
\xi\delta(\rr'-\rr'')
\begin{pmatrix}
\left(\frac{\partial s}{\partial \ww}\right)^2 & \frac{\partial s}{\partial \ww}\\
\frac{\partial s}{\partial \ww} &  \mathbf{1}
\end{pmatrix}.
\end{equation}
Both $\LL$ and $\MM$ operators form the Onsager-Casimir operator \eqref{eq.K} and satisfy OCRR in the above sense. That is, the variables with the same parity ($s$ and $s$, $\ww$ and $\ww$) are coupled through operators that are symmetric with respect to the simultaneous transposition (adjoint operator) and time reversal. Variables with opposite parities ($s$ and $\ww$) are coupled through operators that are antisymmetric in the same sense. Note that the Hamiltonian coupling contributes also to the symmetric coupling ($\ww$-$\ww$) while the dissipative matrix also contributes to the antisymmetric coupling ($s$-$\ww$). In \cite{hco}, the composition of TRT and transposition were called bare and dressed symmetries.

\section{Details of the Chapman-Enskog approximation}\label{sec.chapman}
This Section contains details of the calculations needed in the first Chapman-Enskog approximation from Section \ref{sec.ch1}.
In particular, it contains calculations of the following integrals. The equilibrium value of the $Q_{ij}$ moment is obtained by
\begin{align}
\int d\pp \frac{p_i p_j}{|\pp|} \frac{1}{e^{e^*c|\pp|/k_B}-1} = \frac{4\pi^5 k_B^4}{45(e^*)^4 c^4}\delta_{ij},
\end{align}
where which is calculated by observing that the integral vanishes if $i\neq j$, by transformation to the spherical coordinates, and by expansion of the fraction into a geometric series. 

Integral 
\begin{align}
\int d\pp \frac{p_i p_j}{|\pp|} \frac{1}{e^{e^*c|\pp|/k_B}e^{\mm^*\cdot\pp/k_B}-1} \approx \frac{4\pi^5 k_B^4}{45(e^*)^4 c^4}\delta_{ij}
\end{align}
is approximated by the Taylor expansion $\frac{1}{y e^x-1} \approx \frac{1}{y-1} + \frac{yx}{(y-1)^2}$, where the zero-th order term represents the equilibrium distribution function, $f_R$, and the first-order term disappears after the integration.

Integral 
\begin{align}
\int d\pp \frac{p_i p_j p_k}{|\pp|} \frac{1}{e^{e^*c|\pp|/k_B}e^{(\mm^*)^l p_l/k_B}-1} \approx -\frac{16\pi^5 k_B^4}{225(e^*c)^5}\left(m^*_i\delta_{kj}+m^*_k \delta_{ji} + m^*_j \delta_{ik}\right)
\end{align}
also starts with the above Taylor expansion, where the zero-th order term does not contribute to the overall integral. The first-order term vanishes if an odd power of $p_i$, $p_j$, $p_k$, or $p_l$ is present, which means that the momenta must come either in two distinct pairs or as the fourth power of a particular component of the momentum. Calculation in the spherical coordinates then leads to the above result. 

\end{document}